\documentclass{article}

\usepackage{amssymb,amsmath,verbatim,mathtools,needspace,enumitem,etoolbox,graphicx,physics,microtype,afterpage,bm}

\usepackage[dvipsnames, usenames]{xcolor}

\usepackage{epsfig}
\usepackage{amsfonts}

\usepackage{hyperref}

\usepackage{dcolumn}        % Align table columns on decimal point

\usepackage{amstext}

\date{\today}

\newcommand{\insertplot}[5]{\begin{figure}
 \hfill\hbox to 0.05in{\vbox to #5in{\vfill
 \inputplot{#1}{#4}{#5}}\hfill}
 \hfill\vspace{-.1in}
 \caption{#2}\label{#3}
 \end{figure}}
 \newcommand{\inputplot}[3]{
 \special{ps: plotfile #1}
}
\newcounter{fig}   
 
\newcommand{\ee}{\end{equation}}
\newcommand{\eea}{\end{eqnarray}}

\begin{document}

\title{
\Large{\bf 
Quasinormal modes of rapidly rotating Einstein-Gauss-Bonnet-dilaton black holes
}
}
 \vspace{1.5truecm}

\author{
{\large }%$^{\ddagger}$
{Jose Luis Bl\'azquez-Salcedo}$^{1}$\footnote{jlblaz01@ucm.es}, 
{Fech Scen Khoo}$^{1,2}$\footnote{fech.scen.khoo@uni-oldenburg.de},
{Burkhard Kleihaus}$^{2}$\footnote{b.kleihaus@uni-oldenburg.de},
and
{Jutta Kunz}$^{2}$\footnote{jutta.kunz@uni-oldenburg.de}
\\
\\
$^{1}${\small 
Departamento de F\'isica Te\'orica and IPARCOS, 
Facultad de Ciencias F\'isicas,}
\\ {\small
Universidad Complutense de Madrid,
28040 Madrid, Spain
}
\\
$^{2}${\small  
 Institut f\"ur  Physik, Universit\"at Oldenburg, Postfach 2503,
D-26111 Oldenburg, Germany
} 
}

\maketitle

\begin{abstract}
Quasinormal modes of rapidly rotating black holes are crucial in understanding the ringdown phase after a merger. 
While for Kerr black holes these modes have been known for a long time, their calculation has remained a challenge in alternative theories of gravity.
We obtain the spectrum of quasinormal modes of rapidly rotating black holes in Einstein-Gauss-Bonnet-dilaton theory without resorting to perturbation theory in the coupling constant.
Our approach is based on a spectral decomposition of the linear perturbations of the metric and the scalar field.
The quasinormal modes agree excellently with %
the perturbatively known slow rotation and weak coupling limits. % 
For large coupling, though, the spectrum changes significantly.
\end{abstract}

%\tableofcontents

%%%%%%%%%%%%%%%%%%%%%%%%%%%%%%%%%%%%%%%%%%%%%%%%%%%%%%%%%%%%%%%%%%%%%%%%%%%%%%%%%%%
\section{Introduction}
%%%%%%%%%%%%%%%%%%%%%%%%%%%%%%%%%%%%%%%%%%%%%%%%%%%%%%%%%%%%%%%%%%%%%%%%%%%%%%%%%%%

The coalescence of black holes proceeds via the inspiral of the binary system, the subsequent merger of the two black holes, and then the ringdown of the highly excited newly formed final black hole, as first observed in the detection of gravitational waves by the LIGO collaboration \cite{LIGOScientific:2016aoc}.
The analysis of the gravitational wave signals requires the theoretical modelling of these processes and the construction of the corresponding waveform templates, based on some gravitational theory.

Up to now, General Relativity (GR) is in accordance with all observations (see e.g.\cite{Will:2018bme}).
However, there are reasons to expect that GR will be superseded by some more complete theory of gravity.
This has led to the formulation and study of a large variety of alternative theories of gravity and their predictions in the realms of cosmology and strong gravity  \cite{Faraoni:2010pgm,Berti:2015itd,CANTATA:2021ktz}.

An attractive well-studied alternative theory of gravity is Einstein-Gauss-Bonnet-dilaton (EGBd) theory, that arises as part of an effective action in string theory \cite{Gross:1986mw,Metsaev:1987zx}.
EGBd theory leads to second order equations of motion and does not feature ghosts \cite{Horndeski:1974wa}.
Its action reads
\begin{eqnarray}  
\label{act}
S=\frac{1}{16 \pi}\int d^4x \sqrt{-g} \left[R - \frac{1}{2}
 (\partial_\mu \varphi)^2
 + f(\varphi) R^2_{\rm GB}   \right],
 \label{action}
\end{eqnarray} 
where the dilaton $\varphi$ is coupled to the Gauss-Bonnet (GB) invariant $R^2_{\rm GB} = R_{\mu\nu\rho\sigma} R^{\mu\nu\rho\sigma}- 4 R_{\mu\nu} R^{\mu\nu} + R^2$ with coupling function $f(\varphi)=\alpha e^{-\varphi}$
and GB coupling constant $\alpha$
(and geometrical units $G=c=1$).

Black hole solutions in EGBd theory have been studied since many years \cite{Kanti:1995vq,Torii:1996yi,Guo:2008hf,Pani:2009wy,Kleihaus:2011tg,Pani:2011gy,Ayzenberg:2013wua,Ayzenberg:2014aka,Kleihaus:2014lba,Maselli:2015tta,Kleihaus:2015aje,Blazquez-Salcedo:2016enn,Cunha:2016wzk,Zhang:2017unx, Blazquez-Salcedo:2017txk,Ripley:2019irj,Pierini:2021jxd,Pierini:2022eim}.
But only recently the full inspiral-merger-ringdown waveform has been modelled in EGBd theory, being the first example performed in an alternative theory of gravity \cite{Julie:2024fwy}.
Representing an additional degree of freedom, the presence of the scalar field leads to dipole radiation in the inspiral phase that can markedly reduce the time till merger, depending on the coupling strength.

To model the ringdown phase of the final black hole knowledge of its spectrum of quasinormal modes (QNMs) should be available in the respective alternative gravity theory.
However, up to now, no fully non-perturbative study of the QNMs of rapidly rotating black holes has been performed in any such theory, since the background solutions themselves no longer possess the high degree of symmetry of the Kerr black holes of GR, leading to a large system of coupled partial differential equations (PDEs) to be solved subject to appropriate boundary conditions.

While recently QNMs of rapidly rotating black holes were reported for a GB theory with a linear coupling function $f(\varphi)=\alpha \varphi$, these calculations still use lowest order perturbation theory in their coupling constant both for the background solution and the modes \cite{Chung:2024ira,Chung:2024vaf}.
Here we report the first QNM spectrum of rapidly rotating black holes in an alternative theory of gravity for the strongly coupled case, focusing on EGBd theory Eq.~(\ref{action}).

Our approach is based on a spectral decomposition of the metric and scalar perturbations on the fully non-perturbative background solutions \cite{Kleihaus:2011tg,Kleihaus:2014lba,Kleihaus:2015aje}.
We have obtained the fundamental $l=2$-led and $l=3$-led modes for azimuthal number $M_z=2$, after reproducing the known Kerr black hole spectrum with high accuracy \cite{Blazquez-Salcedo:2023hwg} and gaining further experience with the approach by studying the spectrum of rapidly rotating phantom wormholes \cite{Khoo:2024yeh}.
Our spectrum of EGBd QNMs agrees excellently with all known limits: the Kerr limit \cite{Berti:2009kk}, the static limit \cite{Blazquez-Salcedo:2016enn,Blazquez-Salcedo:2017txk}, and the limit of slow rotation and small coupling \cite{Pierini:2021jxd,Pierini:2022eim}.

%%%%%%%%%%%%%%%%%%%%%%%%%%%%%%%%%%%%%%%%%%%%%%%%%%%%%%%%%%%%%%%%%%
\section{Theoretical setting}
%%%%%%%%%%%%%%%%%%%%%%%%%%%%%%%%%%%%%%%%%%%%%%%%%%%%%%%%%%%%%%%%%%
\label{setup}

The EGBd action (\ref{action}) leads to the following system of generalized Einstein equation
\begin{equation}
G_{\mu\nu} =  T_{\mu\nu} \ , \ \ \ 
\label{eoms}
\end{equation}
where the effective stress-energy tensor
\begin{eqnarray}
T_{\mu\nu} &=&
-\frac{1}{4}g_{\mu\nu}\partial_\rho {\varphi} \partial^\rho {\varphi}
+\frac{1}{2} \partial_\mu {\varphi} \partial_\nu {\varphi} 
-
\frac{1}{2}\left(g_{\rho\mu}g_{\lambda\nu}+g_{\lambda\mu}g_{\rho\nu}\right)
\eta^{\kappa\lambda\alpha\beta}\tilde{R}^{\rho\gamma}_{\alpha\beta}\nabla_\gamma \partial_\kappa f({\varphi}) \ ,
\label{teff}
\end{eqnarray}
with $\tilde{R}^{\rho\gamma}_{\ \ \; \alpha\beta}=\eta^{\rho\gamma\sigma\tau}
R_{\sigma\tau\alpha\beta}$ and $\eta^{\rho\gamma\sigma\tau}= 
\epsilon^{\rho\gamma\sigma\tau}/\sqrt{-g}$,
and dilaton equation %
\begin{eqnarray}
\label{dil-eq}
\nabla^2 {\varphi} -\alpha e^{-{\varphi}}  R^2_{\rm GB}
 =0.
\end{eqnarray}

\begin{figure}[t!]
\begin{center}
\includegraphics[width=8.5cm,angle=0]{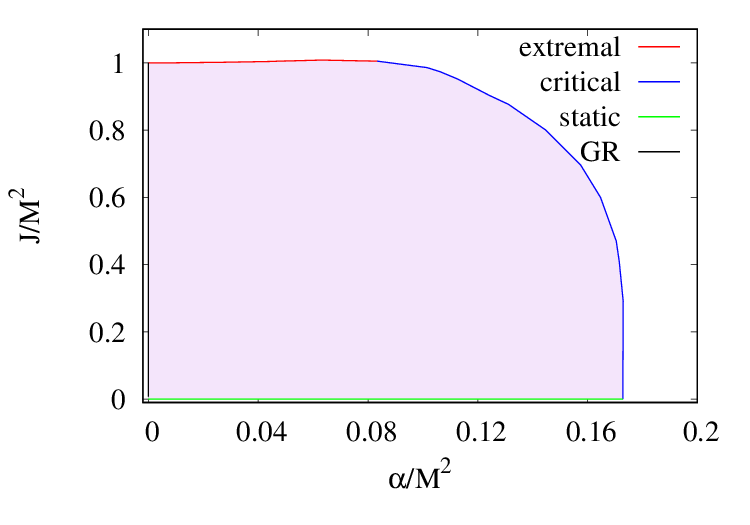} \vspace*{-0.5cm}
\end{center}
\caption{Domain of existence of EGBd black holes:
scaled angular momentum $j=J/M^2$ vs scaled coupling constant $\xi=\alpha/M^2$.
}
\label{fig_bg}
\end{figure}

In Fig.~\ref{fig_bg} we recall the domain of existence of rotating EGBd black hole solutions obtained before \cite{Kleihaus:2011tg,Kleihaus:2014lba,Kleihaus:2015aje}. 
The domain is bounded by the static EGBd black holes \cite{Kanti:1995vq,Torii:1996yi,Guo:2008hf}, the Kerr black holes, the extremal black holes and the critical black holes.
The figure shows the dimensionless angular momentum $j=J/M^2$ (where $M$ is the mass) versus the dimensionless coupling constant $\xi=\alpha/M^2$.

Clearly, these black holes do not exist for arbitrary large coupling, but there is an upper bound on the coupling constant for a given black hole mass $M$, that depends on the angular momentum as well. 
In Fig.~\ref{fig_bg} it can be seen that for values of the angular momentum between $j=0$ and $j=0.286$, the largest value of the coupling is approximately given by $\xi_{max}=0.173$, while for larger values of $j$, the upper bound on $\xi$ is always well below this value.

When employing perturbation theory, care should be taken that for a given $J/M^2$ the boundary line is not too closely approached or even transgressed \cite{Chung:2024ira,Chung:2024vaf}.

To obtain the QNM spectrum we linearly perturb the metric and the dilaton field
\begin{eqnarray}
g_{\mu\nu} &=& g^{(bg)}_{\mu\nu} + \epsilon \delta h_{\mu\nu}(t,r,\theta,\phi) \, , \\
\varphi &=& \varphi^{(bg)} + \epsilon \delta\varphi(t,r,\theta,\phi) \, {,}
\end{eqnarray}
with axial $(A)$ and polar $(P)$ metric components 
$ \delta h_{\mu\nu} = \delta h^{(A)}_{\mu\nu} + \delta h^{(P)}_{\mu\nu} $. 
We then factor the dependence on time $t$ and azimuthal angle $\phi$, introducing the complex eigenfrequency $\omega$ and azimuthal number $M_z$. 

Insertion of these expansions yields the coupled set of perturbation equations
\begin{eqnarray}
\mathcal{G}_{\mu\nu} = \mathcal{G}_{\mu\nu}^{(bg)} + \epsilon \delta\mathcal{G}_{\mu\nu} (r,\theta) e^{i(M_z\phi-\omega t)}  =0 \, , \\
\mathcal{S} = \mathcal{S}^{(bg)} + \epsilon \delta\mathcal{S} (r,\theta) e^{i(M_z\phi-\omega t)}   =0 \, ,
\end{eqnarray}
where the rotating EGBd background solutions satisfy the equations $\mathcal{G}_{\mu\nu}^{(bg)}=0$ and $\mathcal{S}^{(bg)}=0$.
The components $\delta\mathcal{G}_{\mu\nu}$ and $\delta\mathcal{S}$ then result in a system of partial differential equations (PDEs) in $r$ and $\theta$ for a set of seven unknown perturbations functions (for details see \cite{Blazquez-Salcedo:2023hwg,Khoo:2024yeh}).

Next we compactify coordinates, $x = \frac{r-r_H}{r+1}$ and $y = \cos\theta$ with domains of integration $0 \le x \le 1$ and $-1 \le y \le 1$, respectively. 
This places the horizon at $x=0$, asymptotic infinity at $x=1$, and the symmetry axis at $y=\pm 1$. 
The QNMs of the EGBd black holes should be purely ingoing at the horizon and purely outgoing at infinity.
This is taken into account by factorizing the corresponding asymptotic behavior in the perturbation functions leading to a new set of seven unknown functions, that can be collected in a vector $\vec{X}(x,y)$, that should satisfy the resulting system of 7 linear and homogeneous PDEs  
\begin{equation}
    \mathcal{D}_{\mathrm{I}}(x,y) \vec{X}(x,y) = 0, \, \, \, \, \quad   \mathrm{I} = 1,...,7 \, ,
    \label{metric_eq_xy}
\end{equation}
subject to the appropriate set of boundary conditions \cite{Blazquez-Salcedo:2023hwg,Khoo:2024yeh}.

As numerical approach we then employ a spectral method, where we decompose the unknown perturbation functions in a series of Chebyshev polynomials $T_k(x)$ and Legendre functions $P_l^{M_z}{(y)}$, expressing each component $X_I$ of the vector $\vec{X}$ as
\begin{equation}
X_I(x,y) = \sum_{k=0}^{N_x-1} \, \, \sum_{l=|M_z|}^{N_y+|M_z|-1}  C_{I,k,l}  \,  T_k(x)  \,  P_l^{M_z}(y) 
\, .
\end{equation}
The $7 \times N_x \times N_y$ unknown expansion coefficients $C_{I,k,l}$ are constants {to be} determined by solving the system of PDEs subject to the imposed boundary conditions on a grid of $N_x \times N_y$ points{. A} typical grid size has $N_x=N_y=20$, for which all the PDEs are satisfied with an accuracy of at least $10^{-4}$.

The resulting set of equations may be expressed in a matrix form 
\begin{equation}
    \left( \mathcal{M}_0 + \mathcal{M}_1 \omega + \mathcal{M}_2 \omega^2 \right) \Vec{C} = 0 \, ,
    \label{matrix_eq}
\end{equation}
with $(7 \times N_x \times N_y) \times (7 \times N_x \times N_y)$ square matrices $\mathcal{M}_0$, $\mathcal{M}_1$ and $\mathcal{M}_2$ and coefficient vector $\vec{C}$ comprising all constants $C_{I,k,l}$ with $I=1,...,7$. 
We solve this standard quadratic eigenvalue problem for $\omega$ by employing both Maple and Matlab with the Multiprecision Computing Toolbox Advanpix \cite{Advanpix}.
We cross-check our results by evaluating the obtained QNMs and perturbation functions for the set of PDEs not used in the spectral decomposition.
All 11 PDEs are satisfied within an error smaller than $10^{-4}$ at each point.

%%%%%%%%%%%%%%%%%%%%%%%%%%%%%%%%%%%%%%%%%%%%%%%%%%%%%%%%%%%%%%%%%%
\section{Spectrum}
%%%%%%%%%%%%%%%%%%%%%%%%%%%%%%%%%%%%%%%%%%%%%%%%%%%%%%%%%%%%%%%%%%

\begin{figure*}[t!]
\begin{center}
\mbox{ 
\includegraphics[width=1\textwidth,angle=-90]{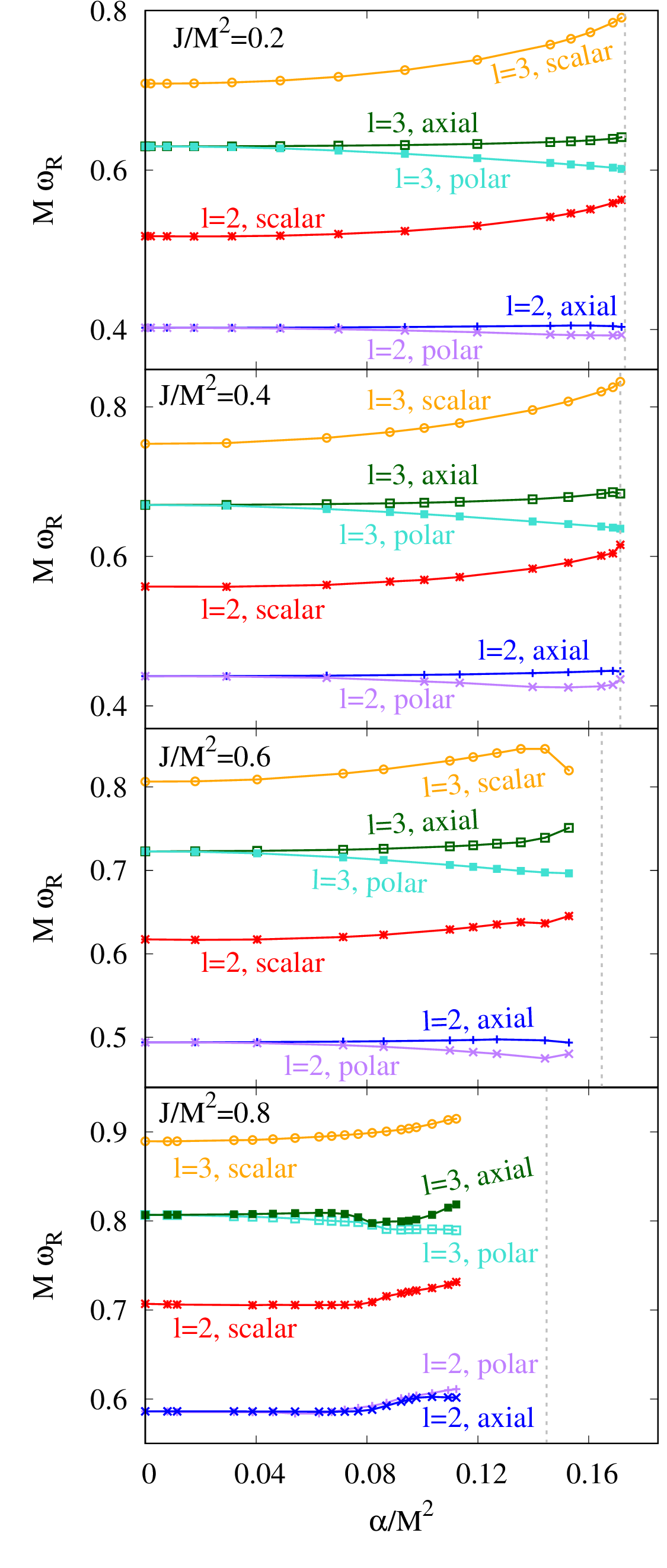}
\includegraphics[width=1\textwidth,angle=-90]{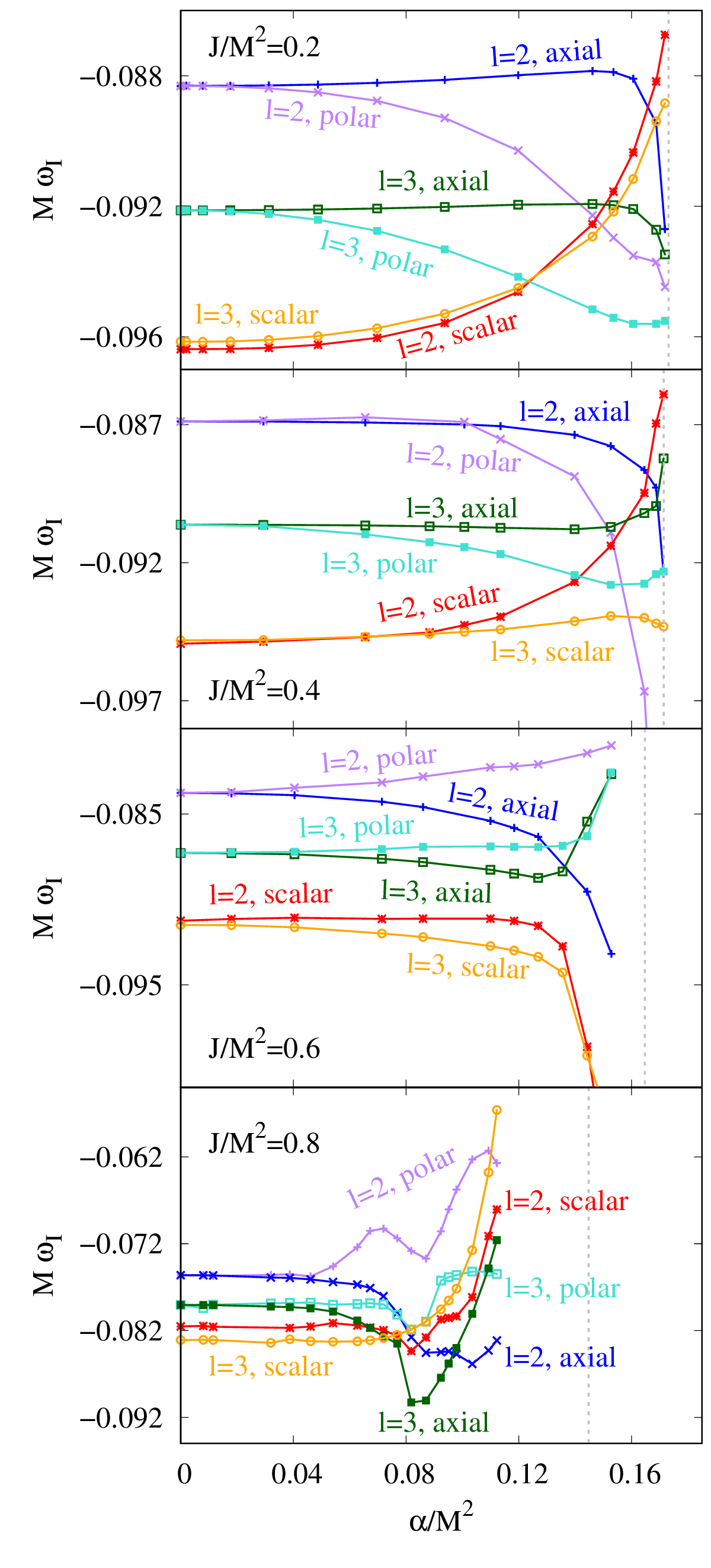}
}
\vspace*{-0.5cm}
\end{center}
\caption{
Fundamental EGBd $(l=2)$-led and $(l=3)$-led QNMs for $M_z=2$:
the scaled real part $M \omega_R$ (left column) and scaled imaginary part $M \omega_I$ (right column) are shown vs the scaled coupling strength $\xi$ for {(from top to bottom) $j=0.2$, $j=0.4$, $j=0.6$, and $j=0.8$} 
with each multipole $l$ featuring distinct polar-led, axial-led and scalar-led modes. {The thin vertical lines indicate the respective maximal couplings.}
}
\label{fig_j0.2}
\end{figure*}

We here concentrate on the lowest fundamental modes, while in principle, the approach also allows for the computation of excited modes \cite{Blazquez-Salcedo:2023hwg,Khoo:2024yeh}.
Also, we restrict to the sector with azimuthal number $M_z=2$, that is expected to be most relevant during the ringdown phase after black hole merger. 
Thus we discuss the fundamental $(l=2)$-led and $(l=3)$-led modes, where the nomenclature $l$-led mode refers to the mode that emerges from the corresponding mode in the static limit.
In our rotating backgrounds, of course, the perturbation functions consist of sums of different $l$-multipoles.
Likewise, axial and polar modes become mixed, as well as the scalar modes.

\begin{figure*}[t!]
\begin{center}
\mbox{ 
\includegraphics[width=6.0cm,angle=-90]{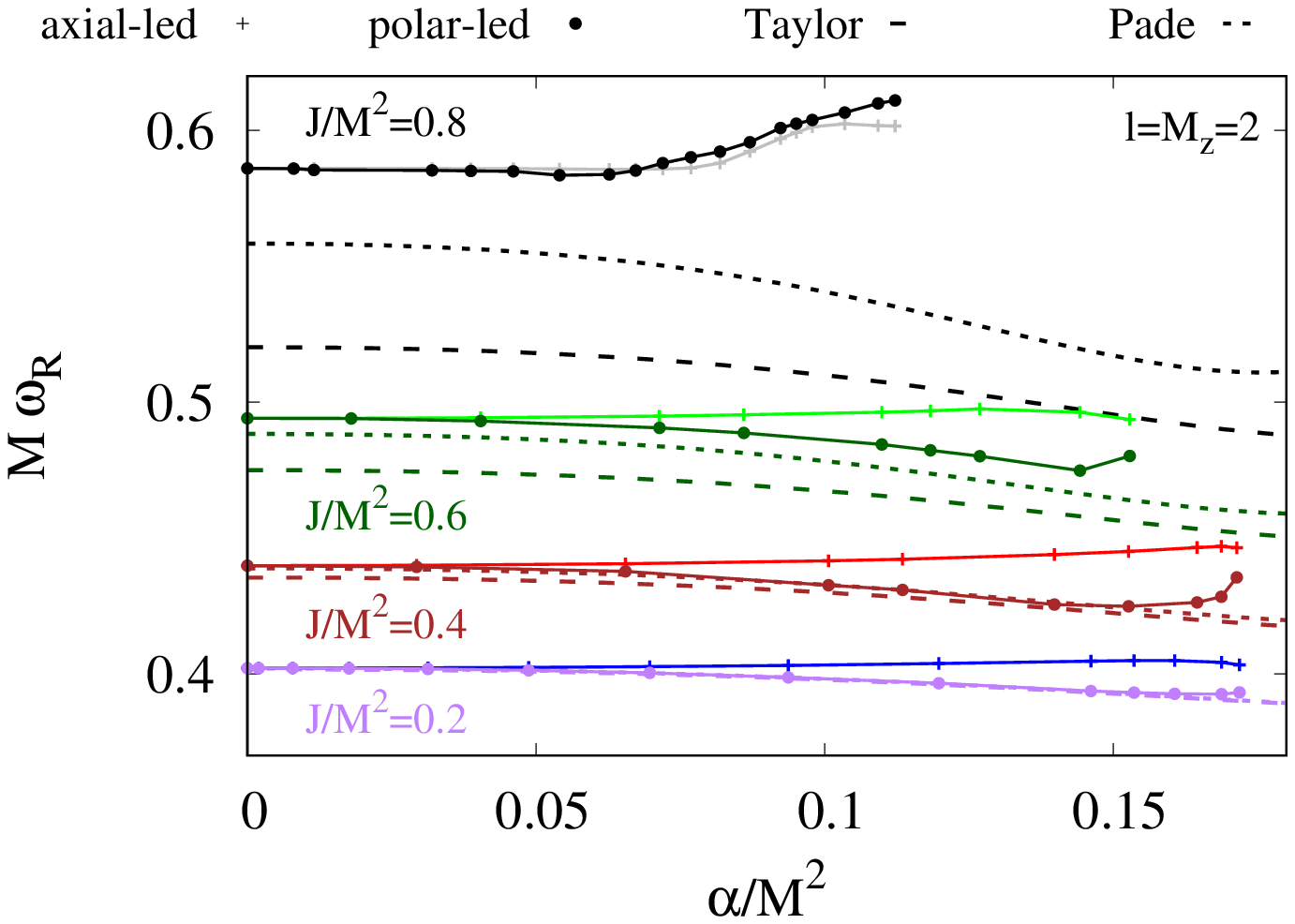}
\includegraphics[width=6.0cm,angle=-90]{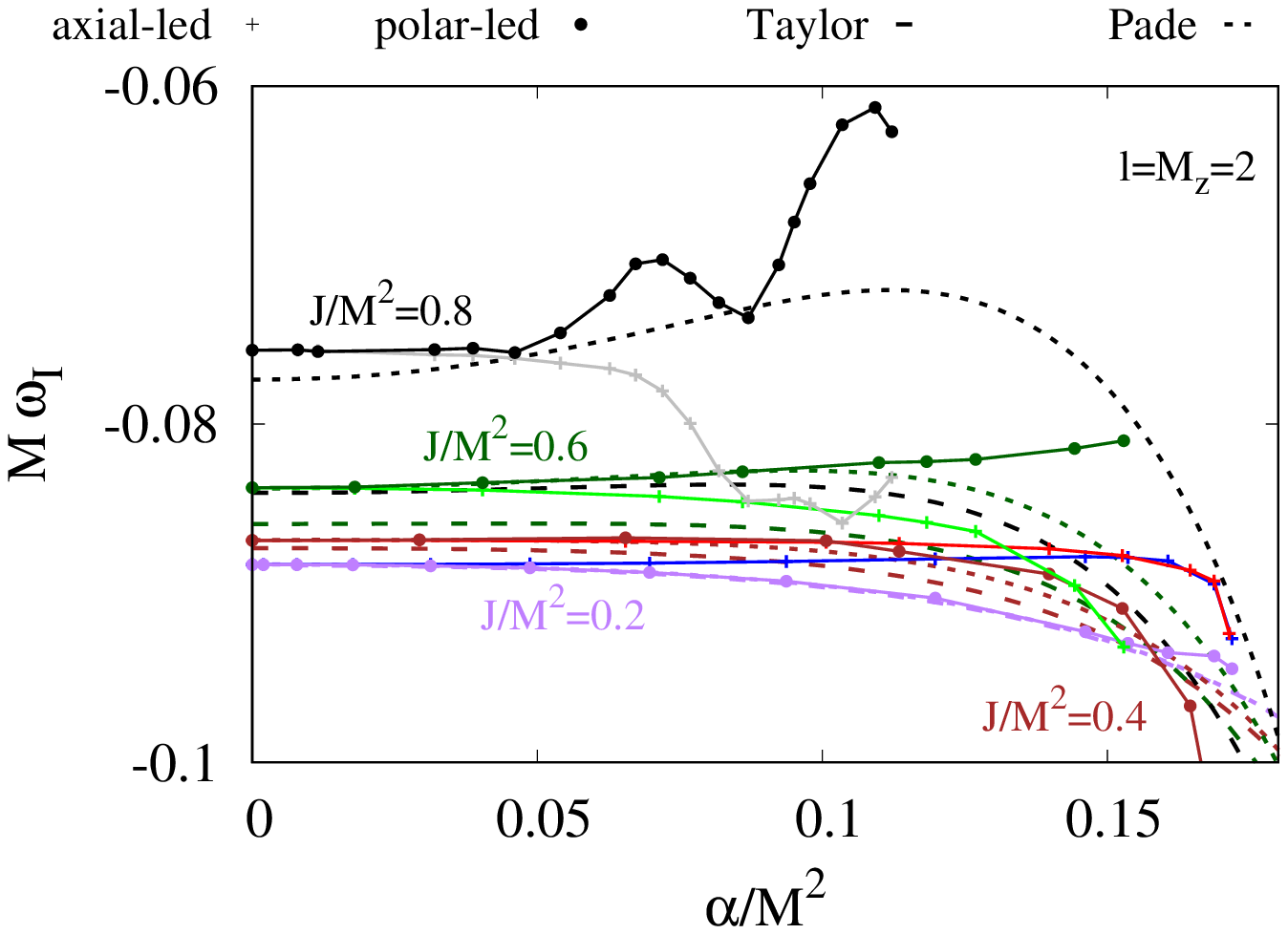}
}
\vspace*{-0.5cm}
\end{center}
\caption{QNM comparison with $M_z=2$, $l=2$-led polar modes obtained in 6th order in coupling strength $\xi$ and 2nd order in rotation parameter $j$ using the {Taylor and the} Pad\'e approximation \cite{Pierini:2022eim}.}
\label{fig_l2m2_vs_approx}
\end{figure*} 

Figure \ref{fig_j0.2} shows the scaled real part $M \omega_R$ (left column) and scaled imaginary part $M \omega_I$ (right column) versus the scaled coupling strength $\xi=\alpha/M^2$ for fundamental $(l=2)$-led and $(l=3)$-led modes with $M_z=2$ and four values of the scaled angular momentum $j=J/M^2$,
$j=0.2,\, 0.4,\, 0.6$ and $0.8$ from top to bottom.
For each multipole $l$ we further distinguish between polar-led, axial-led and scalar-led modes.
In the limit of vanishing coupling strength $\xi$, all branches of EGBd modes smoothly emerge from their corresponding Kerr limits \cite{Berti:2009kk}.

The isospectrality of the Kerr axial-led {and} polar-led modes is broken for the EGBd modes, where the splitting typically increases with increasing coupling strength.
As in the static case, deviations from the limiting GR modes become larger  
towards the boundary of the domain of existence of the background solutions.
The lowest frequency $M \omega_R$ is observed for the $l=2$ polar modes, except for sufficiently large values of the coupling strength $\xi$ and rotation parameter $j$.

In contrast, {the decay time is rather sensitive to the coupling strength and the angular momentum.
Thus} the longest decay time $\tau=1/|\omega_I|$ varies.
For small $j$ the $l=2$ axial mode is dominant for small coupling $\xi$, while close to the maximal $\xi$ the $l=2$ scalar mode starts to dominate.
This may seem surprising at first, however, as $\xi$ increases, the scalar field in the background solution becomes more prominent and this entails a stronger mixing of the gravitational and scalar components in the modes.
For $j=0.4$ the axial and polar $l=2$ decay times remain close to the Kerr values for a large range of $\xi$, but finally the scalar mode dominates again.
For $j=0.6$  the $l=2$ polar mode dominates, while the axial and scalar modes feature much smaller decay times.
For $j=0.8$ the dominance of the modes 
again changes close to the maximal $\xi$.
These results demonstrate, that in order to find the dominant mode, one has to obtain all three types of modes: axial-led, polar-led and scalar-led.

We now compare with previous perturbative results, obtained in 6th order in the coupling strength $\xi$ and 2nd order in the rotation parameter $j$ for the polar modes \cite{Pierini:2022eim}.
We exhibit our axial and polar $l=2$-led modes with $M_z=2$ in Fig.~\ref{fig_l2m2_vs_approx} for angular momenta $j=0.2$, $0.4$, $0.6$
and $0.8$
together with the respective perturbative polar modes
using both the Taylor and the Pad\'e approximation
\cite{Pierini:2022eim}.
Clearly, the Pad\'e approximation captures the change of the modes with increasing angular momentum much better.
We observe excellent agreement in the slowly rotating case $j=0.2$, both for the real part and the imaginary part of the frequency for most of the range of the coupling strength.
Slight deviations arise only in the vicinity of the maximal coupling. 

For rotation parameter $j=0.4$ the agreement is still amazingly good for the real part, while the decay time shows slightly larger deviations.
For even faster rotation ($j=0.6$, {and} {$0.8$}) the deviations grow further, but the general trend is still captured {well} by the perturbative results.
We note, that the increasing size of the deviations is already seen in the Kerr limit, since perturbation theory in $j$ is only second order \cite{Pierini:2022eim}.
Comparison of our $l=3$-led modes with the corresponding perturbative modes shows similar agreement.

Comparison of our results with the recent results obtained in Refs.~\cite{Chung:2024ira,Chung:2024vaf} is hindered by the fact, that those results are obtained for {a linear coupling function}.
They are illustrated for the coupling $\zeta=0.1$ ($\xi= 0.316$),
a value close to the critical boundary of the non-perturbative solutions of the linear theory for small $j$ and outside the domain of existence for large $j$ \cite{Delgado:2020rev}. 
To enable a comparison of exact results and perturbative results we have obtained the exact modes of the 
theory with linear coupling function for $j=0.2$ and $\zeta=0.1$.
Indeed, we find a deviation up to about 5\%
between
our exact modes and the modes of Refs.~\cite{Chung:2024ira,Chung:2024vaf}, {clearly demonstrating that {the perturbative result} is no longer valid for such large coupling.}

%%%%%%%%%%%%%%%%%%%%%%%%%%%%%%%%%%%%%%%%%%%%%%%%%%%
\section{Conclusions}
%%%%%%%%%%%%%%%%%%%%%%%%%%%%%%%%%%%%%%%%%%%%%%%%%%%

The calculation of QNMs of rotating black holes is of utmost relevance to understand the ringdown after merger.
However, in alternative theories of gravity the tools of GR may not work.
Here we have presented the first construction of the QNM spectrum of rapidly rotating black holes in such a theory, by treating the beyond GR terms exactly both on the level of the background solution and on the level of the perturbations.

We have focused on EGBd theory with a dilatonic coupling function, and presented, for the first time, the full calculation of axial, polar and scalar $l=2$-led and $l=3$-led modes for $M_z=2$.
Future calculations will concentrate on the $M_z<2$ cases and on the overtones.
Interestingly, our results show that the dominant mode can be axial, polar or scalar, depending on the rotation parameter $j$ and the coupling strength $\xi$. 
Even more, we have shown that for some range of these parameters, the $l=3$-led modes dominate the $l=2$-led modes.

All this indicates that, strikingly, the postmerger gravitational radiation in theories beyond GR 
could be fundamentally different from that of GR. 
Moreover, all these effects cannot be observed with just a perturbative approximation of the coupling strength. 
But they are of utmost importance, since modelling the ringdown phase with such substantial differences with respect to GR could potentially improve the observational constraints on a theory.

With the new machinery at hand QNMs of rapidly rotating black holes and other compact objects can now be provided for inspiral-merger-ringdown studies \cite{Julie:2024fwy} and comparison with observations.

%%%%%%%%%%%%%%%%%%%%%%%%%%%  
\section*{Acknowledgements}
%%%%%%%%%%%%%%%%%%%%%%%%%%%

We gratefully acknowledge support by DFG project Ku612/18-1, 
FCT project PTDC/FIS-AST/3041/2020, 
and MICINN project PID2021-125617NB-I00 ``QuasiMode''. JLBS gratefully acknowledges support from MICINN project CNS2023-144089 ``Quasinormal modes''. FSK gratefully acknowledges support from ``Atracci\'on de Talento Investigador Cesar Nombela'' of the Comunidad de Madrid, grant no. 2024-T1/COM-31385.

%%%%%%%%%%%%%%%%%%%%%%%%%%%%%%%%%%%%%%%%%%%%%%%%%%%%%%%%%%%%%%%%%%%%%%%%%%%%%%  
\begin{small}

 \end{small}
%%%%%%%%%%%%%%%%%%%%%%%%%%%%%%%%%%%%%%%%%%%%%%%%%%%%%%%%%%%%

\end{document}